\documentclass[apl,showpacs,superscriptaddress,twocolumn,nofootinbib]{revtex4-1}
\usepackage{graphicx}
\usepackage{amsmath}
\usepackage{amssymb}
\usepackage{ulem}
\usepackage{bm}
\usepackage{float}
\usepackage{tikz}
\usepackage[colorlinks,
linkcolor=blue,
anchorcolor=blue,
urlcolor=blue,
citecolor=blue,
]{hyperref}

\begin{document}
	\title{The photon production and collective flows from magnetic induced gluon fusion and splitting in early stage of high energy nuclear collision}
	\date{\today}
	\author{M. R. Jia \footnote{ Email address: jiamr@ccnu.edu.cn}}
	\author{H. X. Li}
	\author{D. F. Hou \footnote{ Email address: houdf@ccnu.edu.cn}}
	\affiliation{Institute of Particle Physics and Key Laboratory of Quark and Lepton Physics (MOS), Central China Normal University, Wuhan 430079, China}
\begin{abstract}
	We present an event-by-event study of photon production in early stage of high energy nuclear collisions, where the system is dominant by highly occupied of gluons and initialized by McLerran-Venugopalan model. The photons are produced through the gluon fusion and splitting processes when strong magnetic field is included. We study the spectra and collective flows of the photons and show their dependence on transverse momentum $q_{T}$. It is found that in our approach the photons from boost invariant evolving glasma provide visible enhancement on spectrum and obvious contribution on $v_{2}$ of the total direct photons. The results, by weighting on top of parton-hadron-string dynamics (PHSD) model, agree even better with experiment measurements in Au-Au 20\%-40\% centrality collisions at $\sqrt{s_{NN}}=200$GeV.
\end{abstract}
\maketitle

\section{Introduction}\label{sec1}
As is known, photons are important probes in heavy-ion collisions, which provide fruitful signals to investigate the pre-equilibrium stage of the collisions, the quark-gluon plasma (QGP), and the hadronic phase. Experiments at Relativistic Heavy Ion Collider (RHIC) and Large Hadron Collider (LHC) have confirmed that elliptic flow of direct photons is large and can be compared to that of hadrons \cite{PHENIX:2011oxq,PHENIX:2015igl,PHENIX:2014nkk,ALICE:2015xmh,ALICE:2018dti}. These measurements have triggered a large amount of theoretical work
on modeling and interpreting the phenomenon \cite{Kapusta:1991qp,Liu:2008eh,Linnyk:2013wma,Ruggieri:2015yea,Paquet:2015lta,Linnyk:2015rco,Linnyk:2015tha,Berges:2017eom,David:2019wpt,Shen:2013cca,Chatterjee:2005de,Liu:2012ax,McLerran:2014hza,Dasgupta:2019whr,Ruggieri:2017efm,Oliva:2017pri,Greif:2017nct,Greif:2016jeb}. Beside all the achievements made by the physicists, theoretical calculations still underestimate the measurements so far. This is the so called "direct photon puzzle".

Event-by-event hydrodynamic model provide a time-dependent environment for photon generations, the state-of-the-art computation has found a reasonable agreement for low $q_{T}$ photon, see \cite{Paquet:2015lta}, the computation shows a significant contribution from late stage photon emission as well as a sizable $v_{2}$. Partonic channels also constitute the part of observed direct photons. The parton-hadron-string dynamics (PHSD) model copes the full evolution of a relativistic heavy-ion collisions, the photon emitting from different QGP stage is systematically studied in \cite{Linnyk:2013wma,Linnyk:2015rco,Linnyk:2015tha}. The PHSD model has greatly achieved in explaining photons from medium, especially, the computation results agree pretty well on $v_2$ and $v_3$ measurement for $q_{T}<2\rm GeV$ region. Meanwhile, to have a full understanding on the photons, other efforts have also been made on modeling the problem, for example other transport models developed in  \cite{Ruggieri:2015yea,Greif:2016jeb,Oliva:2017pri} for early stage, and so on.

As is noticed, strong magnetic field is produced through non-central collisions in early stage of  Relativistic Heavy Ion Collisions \cite{Rafelski:1975rf,SKOKOV_2009,Voronyuk:2011jd,Deng:2012pc,BZDAK2012171,PhysRevC.88.024911}. Thus, all known sources of photon emission experience such magnetic environment. The magnetic field breaks translation invariance, photon emitting from magnetized medium becomes possible candidate for solving the direct photon $v_2$ problem. Models have been developed in various circumstances recently, for example, method through gauge/gravity duality \cite{Muller:2013ila}; photon from conformal anomaly \cite{Basar:2012bp} as well as chiral anomaly \cite{PhysRevD.86.071501}; the photon production by quark splitting and annihilation in strong magnetic field \cite{Wang:2020dsr}.

Noteworthily, in pre-equilibrium stage of heavy ion collisions, where the large-$x$ partons act as static sources of the small-$x$ modes that constitute the Color-Glass Condensate (CGC) fields inside the two Lorentz-contracted colliding nuclei \cite{PhysRevD.49.2233,PhysRevD.49.3352,PhysRevD.50.2225,doi:10.1146/annurev.nucl.010909.083629,GELISCOLOR}. By interacting of CGC fields, the chromoelectric and chromomagnetic fields are formed. This is the glasma \cite{Lappi:2006fp} and the initial stage is gluon-dominated. Thus in such strong magnetized gluonic system, photon can emit through gluon fusion and splitting processes , see \cite{Ayala:2017vex,Ayala:2019jey}. In this work, we present an event by event study on photon emission from a boost invariant evolving glasma. Besides, to avoid the difficulty of a full description for initial stage electromagnetic (EM) fields, we introduce an ansatz of temporal profile to mimic the early evolution of EM fields. Ultimately, we weight our results of the collective flows on top of PHSD model and make comparisons with Au-Au collision measurements at 20\%-40\% centrality in RHIC energy.

We organize the article as follow: In Sec.~\ref{sec2} we briefly review that how to describe the evolving glasma through CYMs with an initial condition provide by McLerran-Venugopalan (MV) model; In Sec.~\ref{sec3} we  recall the photon production in presence of a strong magnetic field through gluon fusions and splittings. Then we deform the problem in a 2+1D boost invariant glasma. In Sec.~\ref{sec4} we do event by event calculation and show our results of the spectra and collective flows. We summarize our computation and make some outlooks on future efforts in Sec.~\ref{sec5}.

\section{The evolving Glasma}\label{sec2}
The dynamics of the central rapidity region is determined by the small Bjorken $x$ gluons before the collision where saturation takes place \cite{Mueller:1999wm,PhysRevD.49.2233,PhysRevD.49.3352,PhysRevD.50.2225}, thus classical Effective Field Theories (EFTs) become useful tools \cite{PhysRevD.55.5414,PhysRevD.54.5463,PhysRevD.59.014015,Iancu_20011,Iancu_20012}. The glasma serves as the initial condition for evolution of the classical gluon field. In early stage, the evolution can be studied by the Classical Yang-Mills (CYM) equations \cite{FUKUSHIMA2012108,PhysRevLett.111.232301,Berges:2020fwq,Gale:2012rq,Gale:2012in,PhysRevD.97.076004} up to formation of the quark-gluon plasma. In this work, the gauge fields have been rescaled by the QCD coupling $A_{\mu}\rightarrow A_{\mu}/g$,  therefore $g$ does not appear explicitly in the equations.

In the MV model the color charge densities $\rho^a$ act as  static sources of the transverse CGC fields in the two colliding nuclei: they are assumed to be random variables that, for each nucleus, are normally distributed with zero average and with variance specified by the equation
\begin{equation}
	\langle\rho^a(\mathbf{x}_{T},\eta_1)\rho^b(\mathbf{y}_{T},\eta_2)\rangle=
	(g^2\mu)^2\delta^{ab}\mathbf{\delta}(\mathbf{x}_{T}-\mathbf{y}_{T})
	\delta(\eta_1-\eta_2),\label{eq2.2.1}
\end{equation}
where $a$ and $b$ denote the adjoint color indices, $\mathbf{x}_{T}$ and $\mathbf{y}_{T}$ denote transverse plane coordinates. $g^2\mu$ above is the only energy scale in the model which is related to the saturation momentum $Q_s$ \cite{LappiWilson}:
we refer to the estimation that $Q_s/g^2\mu\approx 1.15$. To specify the initial condition, it is convenient to work in Bjorken coordinates $(\tau,\eta)$, where
\begin{eqnarray}
	t=&\tau \rm cosh\eta,\label{eq2.2.2}\\
	z=&\tau \rm sinh\eta,\label{eq2.2.3}
\end{eqnarray}
and in the radial gauge, where $A_{\tau}=0$.
In order to compute the glasma fields we firstly solve the Poisson equations, namely
\begin{eqnarray}
	-\mathbf{\nabla}\cdot\alpha^{(A)}(\mathbf{x}_{T})=\rho^{(A)}(\mathbf{x}_{T}),\label{eq2.2.4}\\
	-\mathbf{\nabla}\cdot\alpha^{(B)}(\mathbf{x}_{T})=\rho^{(B)}(\mathbf{x}_{T}),\label{eq2.2.5}
\end{eqnarray}
with $A$ and $B$ denoting the two colliding nuclei. The solutions of these equations are
\begin{eqnarray}
	\alpha^{(A)}_{i}(\mathbf{x}_{T})=iU^{(A)}(x_{T})\partial_iU^{(A)\dagger}(x_{T}),\label{eq2.2.6}\\
	\alpha^{(B)}_{i}(\mathbf{x}_{T})=iU^{(B)}(x_{T})\partial_iU^{(B)\dagger}(x_{T}),\label{eq2.2.7}
\end{eqnarray}
where the Wilson line is defined as $U(x_{T})\equiv\mathcal{P}{\rm exp}\left(-i\int d\mathbf{z}^{\mu}\alpha_{\mu}(\mathbf{z(x_{T})})\right)$, with $\mathcal{P}$ being the path order operator and $\mathbf{z}(x_{T})$ is trajectory. In terms of these fields, the glasma gauge potential at $\tau\rightarrow 0^+$ can be written as \cite{PhysRevD.52.6231,PhysRevD.52.3809}:
\begin{eqnarray}
	&A_{i}=\alpha_{i}^{(A)}+\alpha_{i}^{(B)}, i=x,y,\label{eq2.2.8}\\
	&A_{\eta}=0.\label{eq2.2.9}
\end{eqnarray}
Solving the Yang-Mills equations near the light cone, one finds that the transverse color electric and color magnetic fields vanish as $\tau\rightarrow 0$, but the longitudinal electric and magnetic fields are non-vanishing \cite{Fries_2006}:
\begin{eqnarray}
	&E_{\eta}=i\sum_{i}[\alpha_{i}^{(A)},\alpha_{i}^{(B)}], \label{eq2.2.10}\\
	&B_{\eta}=i([\alpha_{x}^{(A)},\alpha_{y}^{(B)}]+[\alpha_{x}^{(B)},\alpha_{y}^{(A)}]).\label{eq2.2.11}
\end{eqnarray}
In all the discussion above we have neglected the possibility of fluctuations that, among other things,
would break the longitudinal boost invariance. We also assume that $g^2\mu$ has no dependence on the transverse plane coordinates. To mimic the energy density profile that would be produced in realistic collisions \cite{Gale:2012rq,Gale:2012in}, we will remove this assumption in future work.

After preparing the initial condition of CYM equations, within the gauge $A_{\tau}=0$ the Lagrangian density reads
\begin{equation}
	\mathcal{L}={\rm Tr}\big[-\frac{1}{\tau}(\partial_{\tau}A_{\eta})^2-\tau(\partial_{
		\tau}A_{i})^{2}+\frac{1}{\tau}F^{2}_{\eta i}+\frac{\tau}{2} F_{ij}^{2}\big],\label{eq2.2.12}
\end{equation}
and the canonical momenta are defined by
\begin{equation}
	E_{i}=\tau\partial_{\tau}A_{i},\qquad E_{\eta}=\frac{1}{\tau}\partial_{\tau}A_{\eta}.\label{eq2.2.14}
\end{equation}
As a consequence, the Hamiltonian density at mid-rapidity is
\begin{equation}
	\mathcal{H}={\rm Tr}\big[\frac{1}{\tau}E_{i}^2+\tau E_{\eta}^2+\frac{1}{\tau}F_{\eta i}^2+ \frac{\tau}{2}F_{ij}^2\big].\label{eq2.2.15}
\end{equation}
Thus we can identify Hamiltonian by:
\begin{equation}
	H(\tau)=\int dx_{T}^2\mathcal{H}(\mathbf{x}_{T},\tau)=\int\frac{d^2p_{T}}{(2\pi)^2}w(p_{T})n(\tau,p_{T}),
\end{equation}
where $\omega_{p} = |p_{T}|$ is the free dispersion relation of gluons at initial stage, and $n(\tau,p_{T})$ is their occupation number. We use the equation above to estimate the occupation number in evolving glasma and the relation is as follow:
\begin{equation}
	\begin{aligned}
		n(p_{T},\tau)=\frac{1}{|p_{T}|}\rm{tr}\bigg(&\frac{1}{\tau}E_{i}(p_{T},\tau)E_{i}(-p_{T},\tau)\\
		&+\tau E_{\eta}(p_{T},\tau)E_{\eta}(-p_{T},\tau)\\
		&+\frac{1}{\tau}F_{\eta i}(p_{T},\tau)F_{\eta i}(-p_{T},\tau)\\
		&+ \frac{\tau}{2}F_{ij}(p_{T},\tau)F_{ij}(-p_{T},\tau)\bigg).\label{eq17}
	\end{aligned}
\end{equation}

Due to the large occupation number of gluon, we can use classical equation of motion to describe evolution of the system. Then CYM equations in Bjorken coordinates are:
\begin{eqnarray}
	\centering
	&\partial_{\tau}E_{i}=\frac{1}{\tau}\mathcal{D}_{\eta}F_{\eta i}+\tau\mathcal{D}_{j}F_{ji},\label{eq2.2.16}\\
	&\partial_{\tau}E_{\eta}=\frac{1}{\tau}\mathcal{D}_{j}F_{j\eta},\label{eq2.2.17}
\end{eqnarray}
with $\mathcal{D}_{\mu}=\partial_{\mu}+iA_{\mu}$ is the covariant derivative.

\section{Photon emitting from magnetized gluonic system}\label{sec3}

As mentioned in Sec.~\ref{sec1}, the early stage dynamics immediately after the collision can be described by highly occupied gluon system. In non-central collisions, magnetic field is produced intensively at early stage. Hence, in this scenario, photon can be produced by gluon fusion and splitting in a magnetized medium. The fermion propagator in presence of a constant magnetic field, which can be read as in \cite{Schwinger:1951nm}:
\begin{equation}
	S(x,y)=\phi(x,y)\int\frac{d^4p}{(2\pi)^4}e^{-i(x-y)p}\tilde{S}(p)
\end{equation}
in which $\tilde{S}(p)$ is the translational invariant part in momentum space. The total propagator carries a Schwinger phase $\phi(x,y)=e^{ie_{f}\int_{s_{1}}^{s_{2}}dz^{\mu}(s)(A_{\mu}(z)+F_{\mu\nu}(z(s)-z(s_1))^{\nu})}$, where the condition $z(s_{1})=x,z(s_{2})=y$ is satisfied.

In Landau level representation, the translation invariant part of the fermion propagator is also defined as follow \cite{Miransky:2015ava}:
\begin{equation}
	\tilde{S}(p) = e^{-\frac{p_{T}^2}{|e_{f}B|}}\sum_{n=0}^{\infty}\frac{(-1)^{n}\mathcal{D}_{n}(p)}{(p_{0}+i\epsilon)^2 -m^2 -p_{3}^2 - 2n|e_{f}B|}
\end{equation}
where $\mathcal{D}_{n}(p)$ is:
\begin{equation}
	\begin{aligned}
		\mathcal{D}_{n}(p)&\equiv 2(\gamma_{P}\cdot p_{P}+m)\bigg(\mathcal{P}^{+}L_{n}(\frac{2p_{T}^2}{|e_{f}B|})-\\
		&\mathcal{P}^{-}L_{n-1}(\frac{2p_{T}^2}{|e_{f}B|})\bigg) + 4(\gamma_{T}\cdot p_{T})L_{n-1}^{1}(\frac{2p_{T}^2}{|e_{f}B|})
	\end{aligned}
\end{equation}
with $\mathcal{P}^{\pm}=\left[1\pm i\gamma^{1}\gamma^{2}sign(e_{f}B)\right]/2$ is the projection operator, and $L_{n}^{\alpha}(x)$ is the general Laguerre polynomials. Transverse and parallel components are projected by metric with respect to the direction of magnetic field. 

Working in massless limit, we consider that $2|eB|$ is large with respect to squared parallel components of the loop momenta. This is reasonable, since in central rapidity region the difference between squared parallel momenta of the gluons and the photon is small. To study glasma effects in the photon production in initial stage, we simplify the calculation and consider the first nonzero contribution in such scenario: the first nonzero contribution comes from the two propagators on lowest Landau-level (LLL) and one on the first Landau-level (1LL), while contribution from three propagators on LLL vanishes by considering the property of Dirac matrix.
\begin{figure}[t!]
	\centering
	\includegraphics[width=0.85\linewidth]{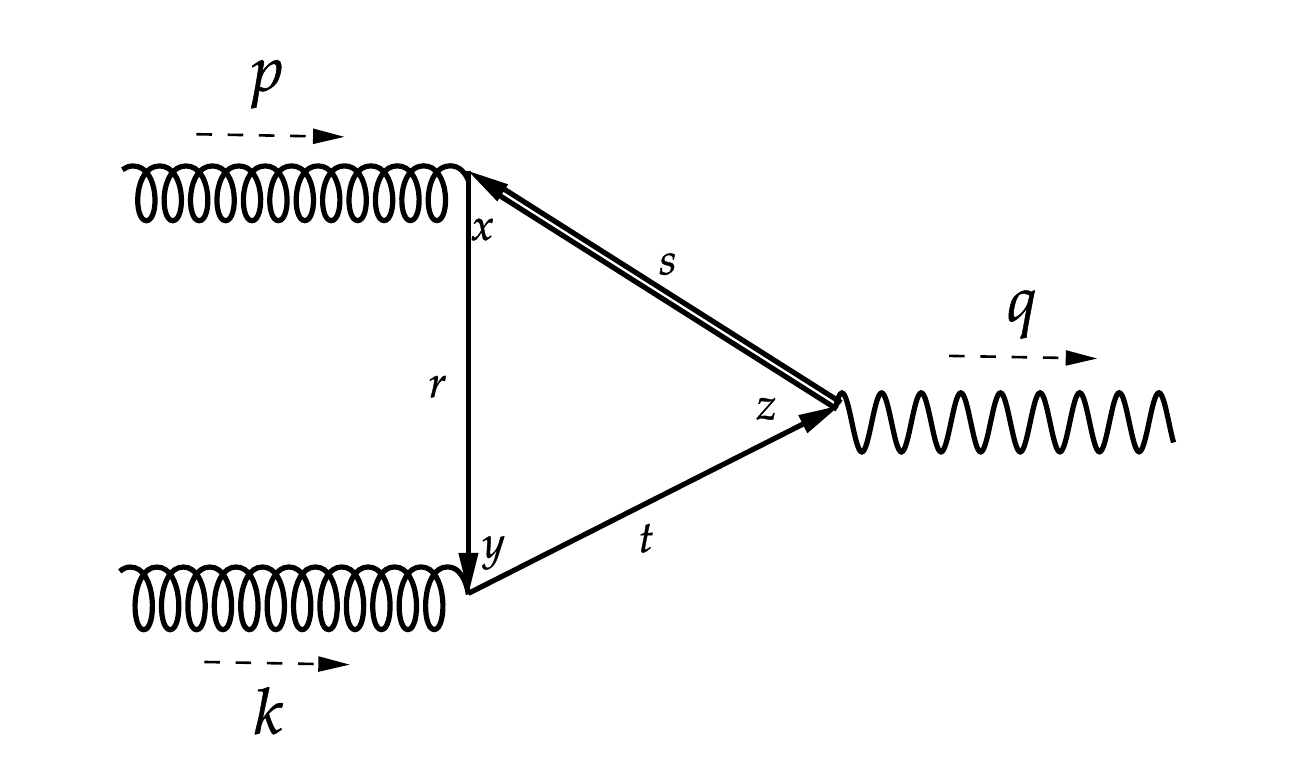}\\
	\includegraphics[width=0.85\linewidth]{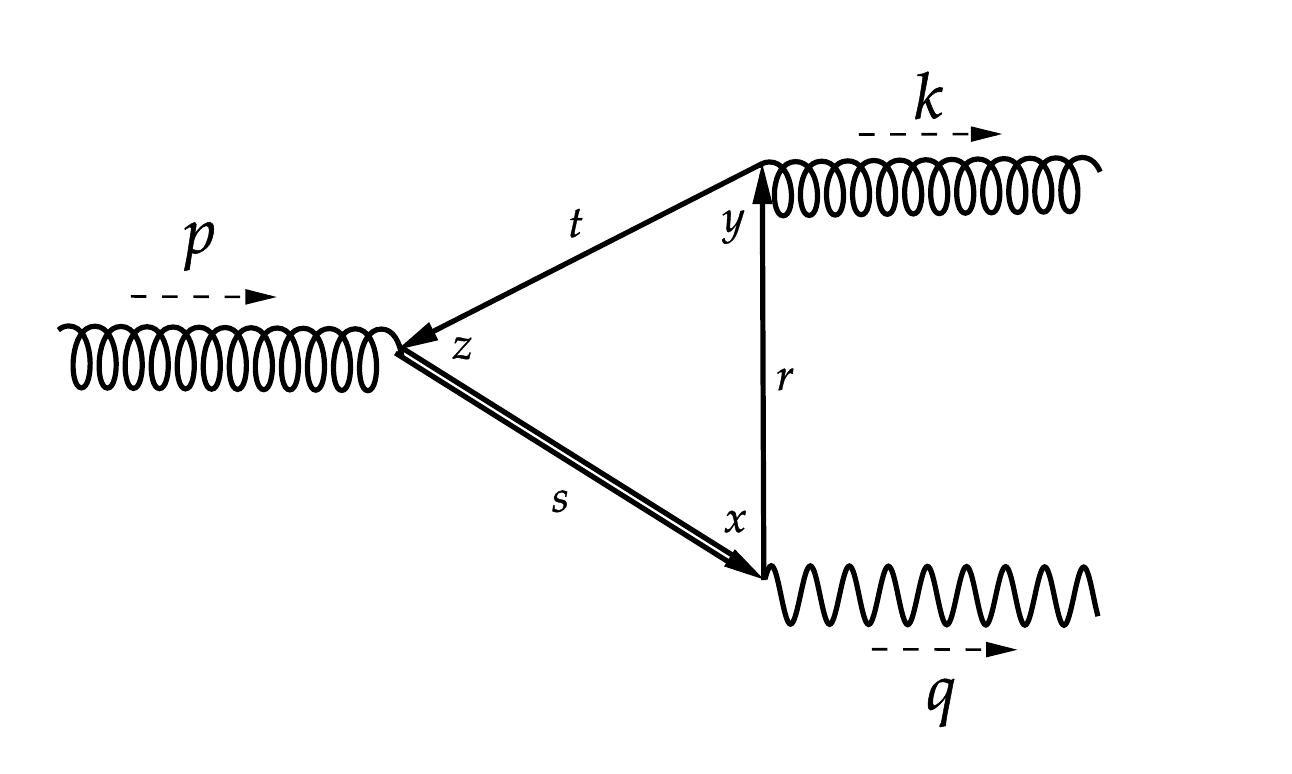}
	\caption{Feynman diagram of photon production in glasma. {\it Upper} : $g+g\rightarrow \gamma$ (gluon fusion); {\it Lower} : $g\rightarrow g\prime + \gamma$ (gluon splitting). Each single line stands for the Lowest Landau Level (LLL) propagator and double line represents the first Landau Level (1LL) propagator; $\rm x, \rm y, \rm z$ denote that each vertex carries a Schwinger phase factor. All non-equivalent permutations of the propagator as well as the charge conjugates are counted when computing the total transition amplitude.}
	\label{fd}
\end{figure}

The perturbative analysis based on Feynman diagram results in the amplitude, see Fig.\ref{fd}:
\begin{equation}
	{\small
		\begin{aligned}
			\mathcal{M}_{g+g\rightarrow \gamma}&=\int d^{4}x\int d^{4}y\int d^{4}z\int\frac{d^{4}r}{(2\pi)^4}\int\frac{d^{4}s}{(2\pi)^4}\int\frac{d^{4}t}{(2\pi)^4}\\
			& e^{-ix(p+t-r)}e^{-iy(r+k-s)}e^{-iz(s-t-q)}\phi(x,y)\phi(y,z)\phi(z,x)\times\\
			&\rm{tr}\left[ie_{f}\gamma^{\alpha}i\tilde{S}_{AB}(s)ig\gamma^{\mu}\rm{t_{BC}^{a}}i\tilde{S}_{CD}(r)ig\gamma^{\mu}\rm{t_{DE}^{b}}i\tilde{S}_{EA}(t) + c.c\right]\\
			&\times\epsilon_{\alpha}(q)\epsilon_{\mu}(p)\epsilon_{\nu}(k),
		\end{aligned}
	}
\end{equation}
while $t_{AB}^{a}$ is the generator in adjoint presentation, and $\gamma^{\mu}$, $\epsilon_{\mu}(p)$ are Dirac matrix and polarization vector respectively. Meanwhile, each vertex carries a Schwinger phase factor. It is similar to process of splittings, $\mathcal{M}_{g\rightarrow g\prime+ \gamma}$. By using above approximation, one can get the averaged amplitude square as in \cite{Ayala:2017vex,Ayala:2019jey}:
\begin{equation}
	\begin{aligned}
		\sum_{p}|\mathcal{M}|^2&=\sum_{p}|\mathcal{M}_{g+g\rightarrow \gamma}|^2=\sum_{p}|\mathcal{M}_{g\rightarrow g\prime+\gamma}|^2\\
		&=\frac{2\alpha_e\alpha_s^2}{N_{c}\pi}\sum_{f}e_{f}^2\left(2\omega_{p}^2+\omega_{k}^2+\omega_{p}\omega_{k}\right)\times\\
		&\frac{q_{T}^2}{\omega_{q}^2}\rm{ exp}\left[-\frac{q_{T}^2}{\omega_{q}^2|e_{f}B|}(\omega_{p}^2+\omega_{k}^2+\omega_{p}\omega_{k})\right],
	\end{aligned}
\end{equation}
while it is averaged over polarization. Here, $\alpha_e$ is the fine structure constant, $\alpha_s$ is the strong coupling and $N_c$ is number of colors and $e_f$ the electric charge number of flavors. Notice that fusion and splitting processes have the same contribution on squared transition amplitude $|\mathcal{M}|^2$. The differential multiplicity, when including both the two effects, is computed as follow:
\begin{equation}
	\begin{aligned}
		\omega_{q}\frac{dN_{\gamma}}{d^{3}p}=\frac{1}{2(2\pi)^3}&\int dx^4\int d\Pi_{p}\int d\Pi_{k}(2\pi)^{4}\\
		&\times\bigg(\delta^{(4)}(q-p-k)	n(\omega_{p})n(\omega_{k})\\
		&+\delta^{(4)}(p-k-q)	n(\omega_{p})(1+n(\omega_{k}))\bigg)\\
		&\times\sum_{p}|\mathcal{M}|^{2}\label{eq2.9}
	\end{aligned}
\end{equation}
where $d\Pi_{p}=\frac{d^3p}{(2\pi)^3\cdot 2\omega_{p}}$ is the Lorentz invariant measure in momentum space.

The amplitude $\mathcal{M}$ is transverse to the plane formed by magnetic field and the propagation diretion, \cite{Ayala:2017vex}. In boost invariant computation, we project the problem on the transverse plane of the colliding system: we choose magnetic field lies in $y$ direction and neglect the expansion of the system on transverse plane, thus, we can factor out $\mathcal{S}_{T} = \int d^2 x_{T}$ as transverse overlap area of two colliding nuclei. By considering the energy-momentum conservation on light cone, the polarization direction of the produced photon is paralleled to $q_{x}$. One thing need to mention is that, $q_{x}$ is the only possible polarization direction in our computation, thus $q_{x}=0$ indicates that no such photon yields, and $\omega_{q}=0$ meets the IR singularity. In the next section, we will use a soft cut off to remove the IR singularity, the similar method has been widely applied in studies such as \cite{Lappi:2006hq,LappiWilson,Schenke:2012wb,FUKUSHIMA2012108,Schenke:2012hg}. Then we carry out the event-by-event calculation on $g+g\rightarrow\gamma$ as well as $g\rightarrow g\prime+\gamma$ processes with intense magnetic field. Especially, we explain the collective flow behavior qualitatively by considering the introduced regulator in our approach, and show the improvements on PHSD prediction toward experimental measures with our full numerical computation.

\section{Numerical Results}\label{sec4}	
In this section we present our numerical results. Firstly, we consider a simplification on magnetic fields generated at initial stage. Then, we study the photon spectra and collective flows in Au-Au collision at RHIC energy.

\subsection{An ansatz of electromagnetic field at initial stage }\label{sec4.1}

\begin{figure}[t!]
	\centering
	\includegraphics[width=\linewidth]{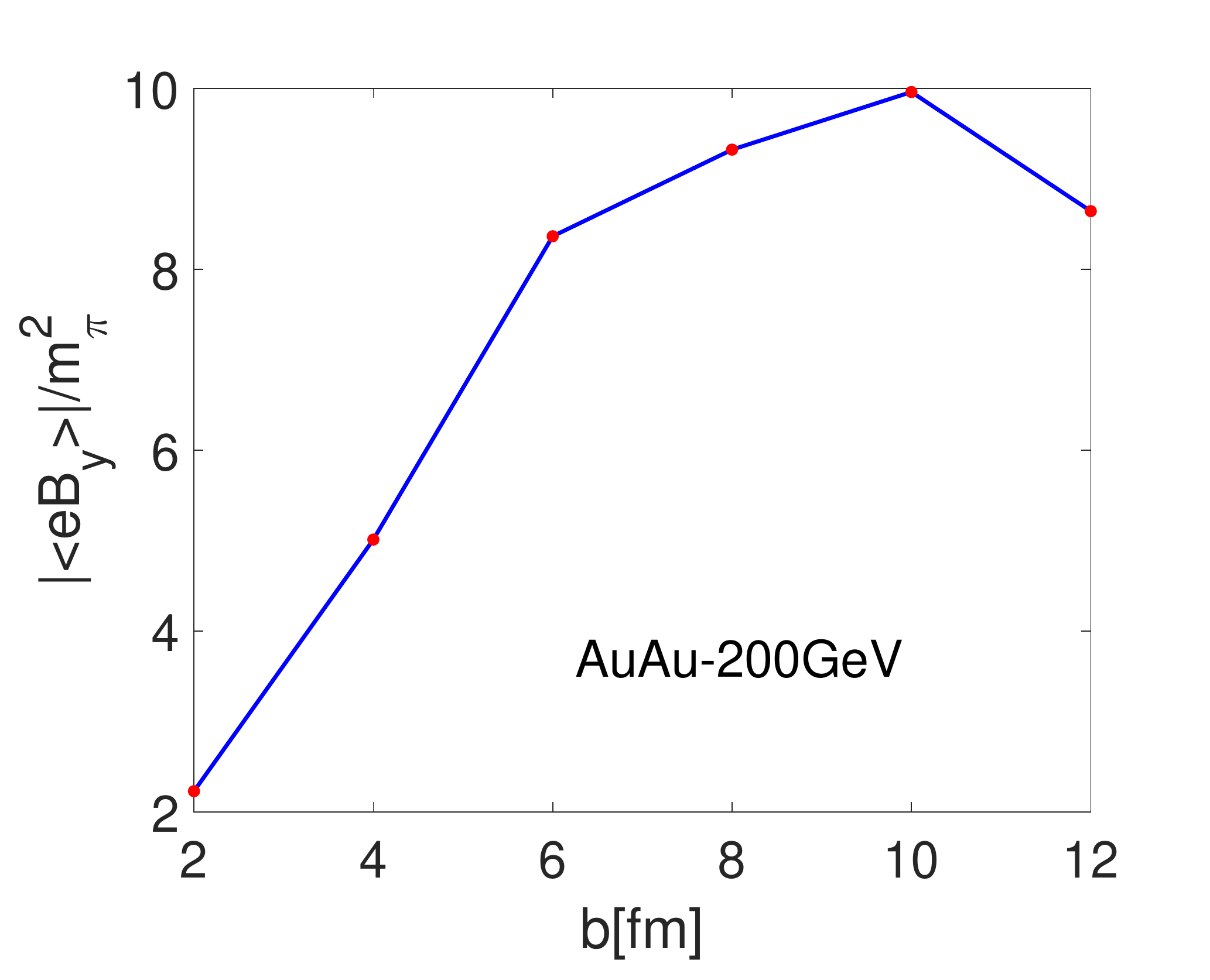}
	\caption{Components of the initial magnetic field at different impact parameters. Au-Au collision at {$\sqrt{S_{NN}}=\rm 200GeV$}. With all parameters can be found in Sec. \ref{sec4.1}.}
	\label{eBb}
\end{figure}

Currently, we focus on overlap region of colliding nuclei, the homogeneously distributed EM fields can be a good approximation. Moreover, we can simplify the EM fields according to the dominant component by
\begin{equation}
	\vec{B}\approx B_y \,, \quad \vec{E}\approx 0 \,.
\end{equation}
To avoid the complexity of initial EM evolution, we introduce a decay function in time for the magnetic field. With the simplifications above, it becomes $\vec{B}(\tau)=(0, B_0f(\tau), 0)$. The decay function is defined by:
\begin{equation}
	f(\tau) = \frac{1}{1+\tau^2/\tau_B^2},
\end{equation}
where $\tau_B$ is the lifetime of magnetic field and acts as a parameter in this work. Here, $B_0 = |\left\langle B_y\right\rangle|$, it's the absolute averaged value in overlap region of two colliding nuclei at $\tau=0$. The average is done as follow: we reproduce the initial fields by using the Li\'enard-Wiechert Potentials:
\begin{equation}
	\begin{aligned}
		&e\vec{E}(t,\bm x)=\alpha_e\sum_{n}\frac{1-v^2}{R^3(1-[\vec{R}\times\vec{v}]^2/R^2)^{3/2}}\vec{R},\\
		&e\vec{B}(t,\bm x)=\alpha_e\sum_{n}\frac{1-v^2}{R^3(1-[\vec{R}\times\vec{v}]^2/R^2)^{3/2}}\vec{v}\times\vec{R}.\label{eq3.1}
	\end{aligned}
\end{equation}
while protons are randomly located according to Woods-Saxon distribution
\begin{equation}
	\rho(r)=\rho_{0}(1+\omega\frac{r^2}{r_0^2})(1+e^{\frac{r-r_0}{a}})^{-1}.\label{eq3.2}
\end{equation}
The method applied in this work is the same as that in \cite{BZDAK2012171}, while all parameters in Eq.(\ref{eq3.2}) can be found in \cite{alver2008phobos}. Moreover, we also do average on  $\mathcal{S}_{T}$, the effective overlap transverse region. This is in agreement with the result from \cite{Deng:2012pc}, while the difference comes from the application of infinitely thin nuclei model as in \cite{BZDAK2012171} as well as the fluctuations. The averaged amplitude of magnetic field at different impact parameters are plotted in Fig.\ref{eBb}.

\subsection{Differential multiplicity of the photons from evolving glasma}\label{p5a}
In the event by event study, we define event average of $\mathcal{O}$ by:
\begin{equation}
	\langle\mathcal{O}\rangle = \frac{1}{N_{event}}\sum_{event}\mathcal{O}_{event}
\end{equation}
For each event, photon from gluon fusion and splitting is computed as follow:
\begin{equation}
	\begin{aligned}
		\omega_{q}\frac{dN_{\gamma}}{d^{3}q}&=\frac{S_{T}\alpha_e\alpha_s^2}{2(2\pi)^5N_c}\sum_{f}e_{f}^2\int_{0}^{\tau}\tau\prime d\tau\prime\int d\omega_{k} \frac{\omega_{p}}{\omega_{k}}\\
		&\times\bigg(n(\omega_{p},\tau\prime)n(\omega_{k},\tau\prime)\delta(\omega_{q}-\omega_{k}-\omega_{p})+(n(\omega_{p},\tau\prime)\\
		&\times(1+n(\omega_{k},\tau\prime))\delta(\omega_{p}-\omega_{k}-\omega_{q})\bigg)
		\left(2\omega_{p}^2+\omega_{k}^2+\omega_{p}\omega_{k}\right)\\
		&\times \frac{q_{x}^2}{\omega_{q}^2}\rm{ exp}\left[-\frac{q_{x}^2}{\omega_{q}^2}\frac{\omega_{p}^2+\omega_{k}^2+\omega_{p}\omega_{k}}{|e_{f}B(\tau\prime)|}\right],\label{eqdm}
	\end{aligned}
\end{equation}
with $S_{T}$ is the transverse area at different impact parameters, we set $N_c=3$ and $e_{f}^2=\{4/9,1/9,1/9\}$ for three flavors of light quarks. There is an IR singularity above in the integral over gluon energy, and it is removed by a soft cutoff, $\Lambda=150\rm MeV$, which equivalently by dressing the gluons with a small mass. The corresponding event average is written as:
\begin{equation}
	\begin{aligned}
		\left\langle\omega_{q}\frac{dN_{\gamma}}{d^{3}q}\right\rangle =\frac{S_{T}\alpha_e\alpha_s^2}{2(2\pi)^5N_c}\sum_{f}e_{f}^2\bigg(\left\langle\int_{0}^{\tau}\tau\prime d\tau\prime f_{g+g\rightarrow\gamma}\right\rangle\\
		+\left\langle \int_{0}^{\tau}\tau\prime d\tau\prime f_{g\rightarrow g\prime+\gamma}\right\rangle\bigg),
	\end{aligned}
\end{equation}
with $f_{g+g\rightarrow\gamma}$ and $f_{g\rightarrow g\prime+\gamma}$ are momentum space integral  part for fusion and splitting in Eq.(\ref{eqdm}) respectively.
\begin{figure}[t!]
	\centering
	\includegraphics[width=\linewidth]{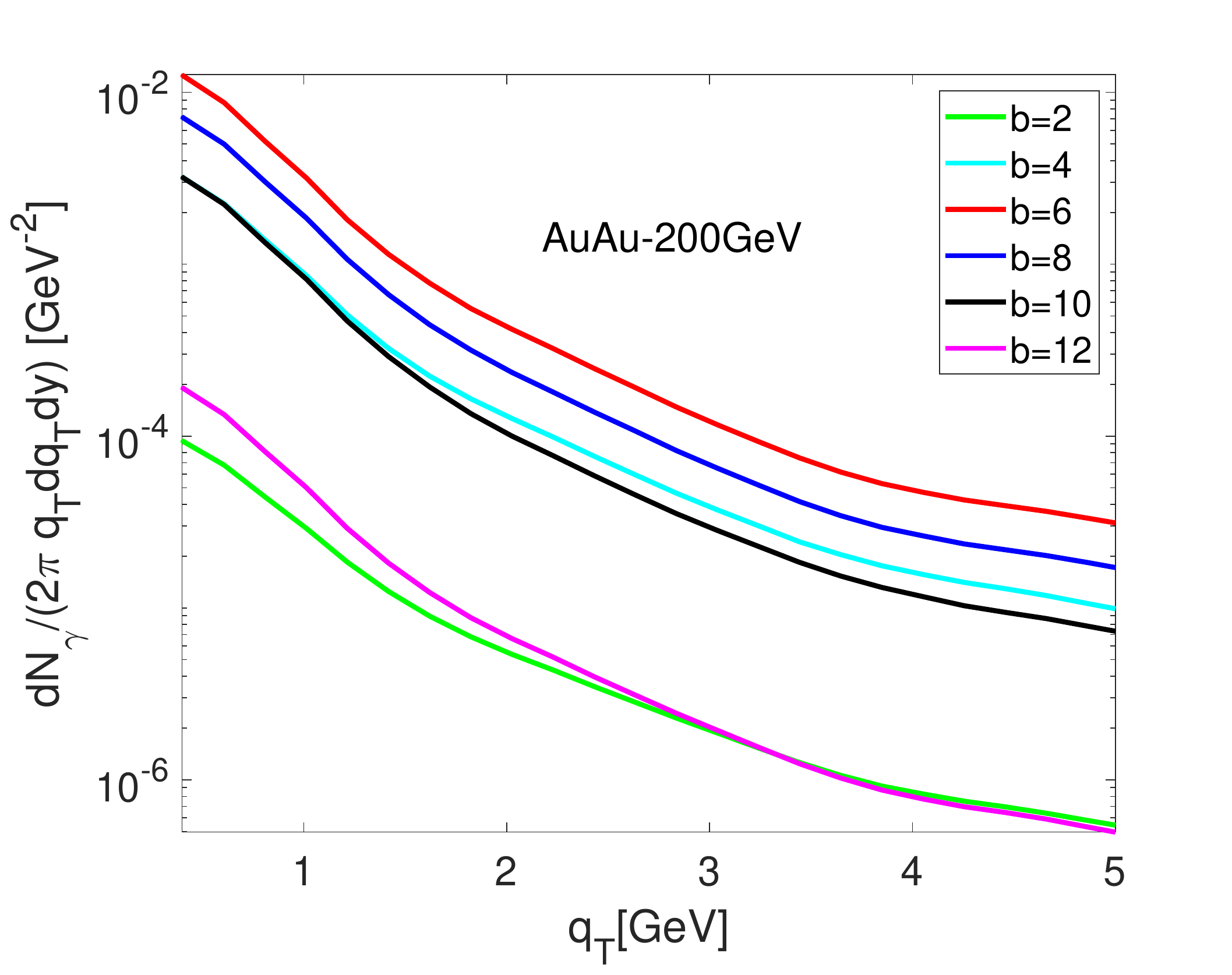}
	\caption{Differential multiplicity of the photons from different centrality cases in evolving glasma up to $0.5\rm{fm/c}$.  Au-Au at $\rm 200GeV$ with $g^2\mu= 1\rm{GeV}$ and $\tau_B = 0.05\rm{fm/c}$, $\alpha_s=1/4\pi$. The lattice spacing is $a=0.03\rm fm$ and the results are averaged over 100 events.}
	\label{dms}
\end{figure}

In each event, gluon occupation number $n(\omega_{p},\tau)$ is computed by Eq.(\ref{eq17}). After integrating over momentum space as well as time, we plot the differential multiplicity of the photons in Fig.\ref{dms}. We use result at different impact parameters to denote spectrum from different centrality classes. Immediately, we can study from Eq.(\ref{eqdm}) that stronger magnetic field can activate higher energy gluons, thus the photon production prefers stronger magnetic field; But increase the strength of magnetic field will not lead to a persistent enhance of the spectrum, since higher energy gluons have lower occupation numbers. Moreover, increasing the impact parameter results in reducing the effective transverse overlap area of two colliding nuclei, thus less participants lead to less number of photons emitting from the medium. Therefore, the photon yield is sensitive to  the centrality classes, depend on two competition effects between the magnetic field and the number of participants. The interplay of these competition effects lead to nonmonotonic behavior of the differential multiplicity of the photons with respect to the impact paremeters shown  in Fig.\ref{dms}.

\begin{figure}[t!]
	\centering
	\includegraphics[width=\linewidth]{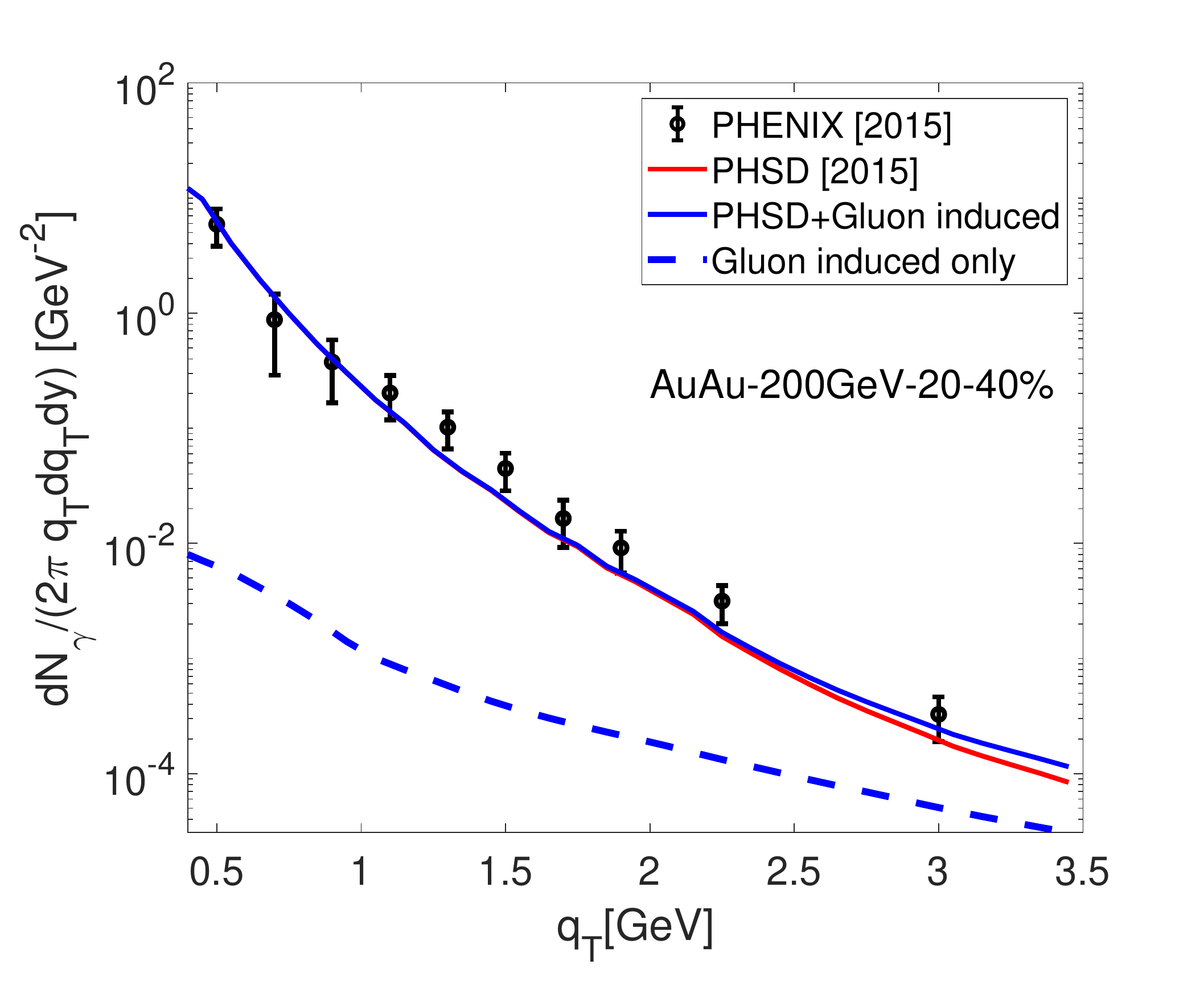}
	\caption{Differential multiplicity as a function of $q_{T}$. The black circles with error bar denote the experiment measures presented in \cite{PHENIX:2014nkk}; Red solid line corresponds to PHSD computation result \cite{Linnyk:2015tha}; The blue solid line is the additive result; The blue dashed line correspond to spectrum only comes from gluon induced processes. The $20\%-40\%$ centrality case is approached at $b=8\rm fm$. All other parameters keep the same as in Fig.\ref{dms}.}
	\label{cdms}
\end{figure}

To make a comparison with results from PHSD model and the experiment measurement, we also plot Fig.\ref{cdms} with $\tau_{B}=$0.05fm/c, where we use result at impact parameter $b=\rm 8fm$ as an approximation to denote the $20\%-40\%$ central collision. It shows that, in $q_{T}< 3\rm GeV$ region, the gluon induced photon spectrum becomes lower as $q_{T}$ goes to zero when compare to PHSD result; while in $q_{T}> 3\rm GeV$, the spectrum becomes comparable to PHSD result as well as experimental measurements. We add the yields on top of PHSD calculation and find only in larger $q_{T}$ part, contribution on total photon spectrum becomes visible. The improvement on PHSD result is still agreed to the measurements and becomes more visualized in $q_{T}> 3\rm GeV$ region. In next subsection, we show that these photons provide obvious contribution on the total elliptic flow of direct photon in larger $q_{T}$ region, and they improve previous PHSD results toward the measurements.

\subsection{Collective flows of the photon}\label{sec4.3}
Expanding the differential multiplicity in Fourier modes:
\begin{equation}
	\omega_{q}\frac{dN_{\gamma}}{dq^3}=\frac{1}{2\pi}\frac{dN}{q_{T}dq_{T}dy}\biggl(1+2\sum_{n=1}^{\infty}v_n\cos[n(\phi-\psi_{RP})]\biggr),\label{eq4.2}
\end{equation}
with $y$ the momentum rapidity, $\phi$ is azimuthal angle, $\psi_{RP}$ is the reaction plane angle (here, we set $\psi_{RP} = 0$). To calculate the coefficient, we can use orthogonality of the trigonometric basis. The event average over $v_{n}$ is computed as follow:
\begin{equation}
	\left\langle v_n(q_{T}) \right\rangle =\int_{0}^{\tau}\tau\prime d\tau\prime\left\langle \frac{\int_{0}^{2\pi}
		d\phi \cos n\phi \big(f_{g+g\rightarrow \gamma}+f_{g\rightarrow g\prime+\gamma}\big)}{2\int_{0}^{2\pi}
		d\phi \big(f_{g+g\rightarrow \gamma}+f_{g\rightarrow g\prime+\gamma}\big)}\right\rangle\label{eq4-3-2}
\end{equation}
while $f_{g+g\rightarrow \gamma}$ and $f_{g\rightarrow g\prime+\gamma}$ have $\phi$ dependence implicitly in equation above. In order to estimate the contribution of these photons on total collective flows, we weight our result on top of PHSD model \cite{Linnyk:2015tha}, and then, compare it to experiment measurements in \cite{PHENIX:2015igl}. The weighted average is defined below:
\begin{equation}
	v_i=\frac{\frac{dN^{\rm PHSD}}{q_{T}dq_{T}dy} v_{i}^{\rm PHSD}+\langle\frac{dN}{q_{T}dq_{T}dy}\rangle\langle v_{i}\rangle}{\frac{dN^{\rm PHSD}}{q_{T}dq_{T}dy}+\langle\frac{dN}{q_{T}dq_{T}dy}\rangle},\label{eq4-3-3}
\end{equation}
here $v_i$ is the corresponding weighted Fourier coefficient, and we compute the cases with $i=2,3$.
\begin{figure}[t!]
	\centering
	\includegraphics[width=\linewidth]{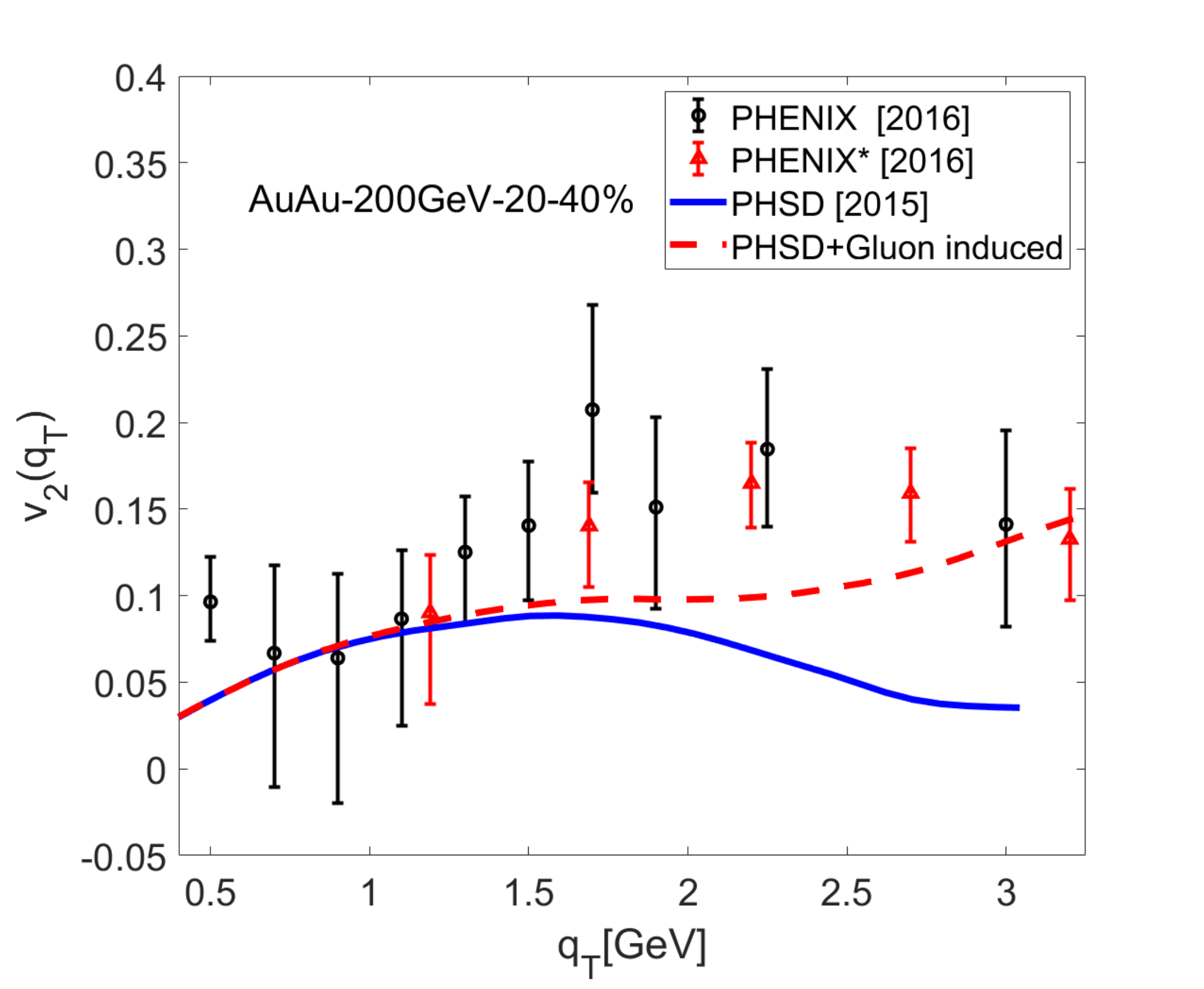}
	\caption{$v_2$ as a function of $q_{T}$. The black circles (Conversion method) and red triangles (Calorimeter method) with error bars denote the experimental measurements presented in \cite{PHENIX:2015igl}; Blue solid line corresponds to PHSD computation result in \cite{Linnyk:2015tha}; The red dashed line stands for weighted result with respect to PHSD mode. All parameters keep the same as in Fig.\ref{cdms}.}
	\label{2}
\end{figure}
\begin{figure}[t!]
	\centering
	\includegraphics[width=\linewidth]{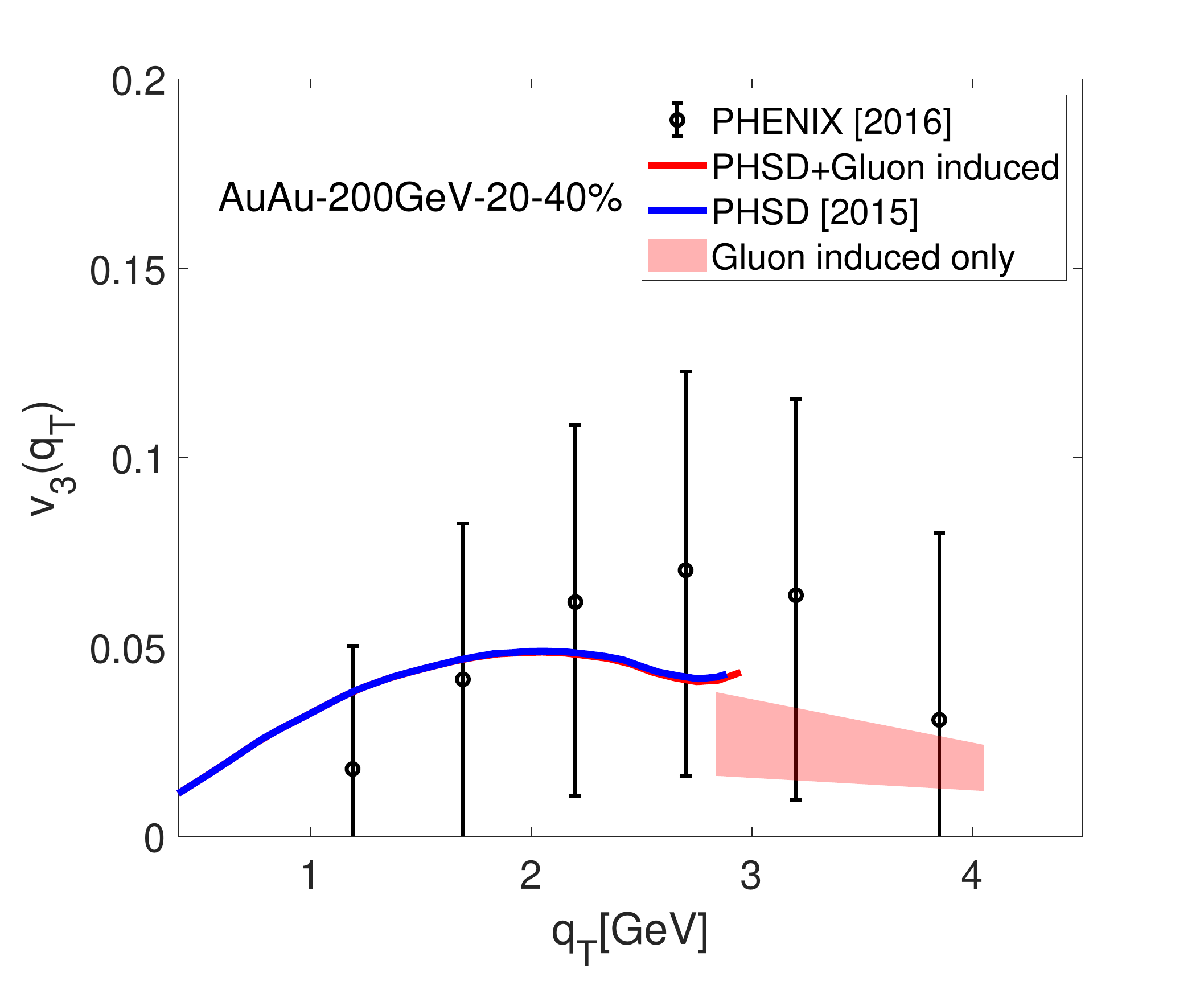}
	\caption{$v_3$ of the photons as a function of $q_{T}$. The black circles with error bars denote the experimental measurements presented in \cite{PHENIX:2015igl} with Calorimeter method; Blue solid line corresponds to PHSD computation result in \cite{Linnyk:2015tha}; The red solid line is the weighted results on top of PHSD model; And colored band comes from the gluon induced processes only. And all parameters keep the same as in Fig.\ref{cdms}.}
	\label{3}
\end{figure}

Before showing the fully numerical results, we can have a qualitatively analysis on behavior of collective flows. Using the free dispersion relation, $\omega_{q}=|q_{T}|$, we can identify $q_{x}/\omega_{q}$ to $\cos\phi$. When considering the regulators in our computation, the angular integral in Eq.(\ref{eq4-3-2}) does not result in modified Bessel functions on lattice. Thus, we factor out the anisotropy part in Eq.(\ref{eqdm}):
\begin{equation}
	\cos^2\phi\rm e^{-\cos^2\phi A}=\frac{q_{x}^2}{\omega_{q}^2}\rm e^{-\frac{q_{x}^2}{\omega_{q}^2}\frac{\omega_{p}^2+\omega_{k}^2+\omega_{p}\omega_{k}}{|e_{f}B|}},
\end{equation}
and carry out an analysis on asymptotic behavior of equation above, while other parts do not carry anisotropy after event average. 

Here, we name $A=(\omega_{p}^2+\omega_{k}^2+\omega_{p}\omega_{k})/|e_{f}B|$, for both cases in fusion and splitting processes. For the region that $(\omega_{p}^2+\omega_{k}^2+\omega_{p}\omega_{k})\gg |e_{f}B|$, $\cos^2\phi$ in exponential is highly smeared by the factor $A$. Thus the dominant term contributing to collective flow becomes $\cos^2\phi$, and this leads to: $v_2\approx1/2$ and $v_{n\neq 2}\approx0$ in high $q_T$ region. In the region that $(\omega_{p}^2+\omega_{k}^2+\omega_{p}\omega_{k})\ll |e_{f}B|$, we can do Taylor expansion on $\cos^2\phi\rm e^{-\cos^2\phi A}$ with respect to $A$. This expansion gives infinite terms carrying nonzero Fourier modes. The linear term of $A$ carries coefficient $\cos^4\phi$, and this provides the first nonzero $\cos3\phi$ term. Consequently, it is the main finite $v_3$ source in our computation.

The fully numerical result agrees with our qualitatively interpretations. To figure out the contribution on total collective flows, we weight our results on top of PHSD model by means of Eq.(\ref{eq4-3-3}), and make comparisons with experiment measurements. The weighted results are plotted in Fig.\ref{2} and Fig.\ref{3}. In Fig.\ref{2}, we have extrapolated the result of \cite{Linnyk:2015tha} from $3\rm GeV$ up to $3.3\rm GeV$ by interpolation when comparing to the measurements in \cite{PHENIX:2015igl}, the small range of interpolation is harmless to the $v_2$ prediction in \cite{Linnyk:2015tha}. In Fig.\ref{2}, we have found that for $q_{T}<\rm 1.5GeV$ region, our result do not destroy PHSD computation, that's because for very soft part the spectrum of photon from gluons is much lower than PHSD estimations. But in $q_{T}>\rm 1.5GeV$ region, the spectrum becomes sizable and the weighted result agrees even better to the experiment measurement on $v_2$ in $1.5\rm GeV< q_{T}< 3.3\rm GeV$ region. Meanwhile in Fig.\ref{3}, $v_3$ of the photon does not affect PHSD result at all. We also attach our result in $3\rm GeV< q_{T}< 4\rm GeV$, where PHSD computation is missing, and the colored band stands for the upper and lower boundary of damping $v_{3}(q_T)$. It is interesting to find that our computation provide the right tendency when compare to the measurement, and this phenomenon has not been shown anywhere else.

\section{Conclusion and Outlook}\label{sec5}

In this work we present an event based study on the
photons emitted through gluon fusion and splitting processes in strong magnetic field in an evolving glasma. The initial condition
is generated by MV model and the system is evolved by CYMs. We use the boost invariant assumption to dress the highly anisotropic property of initial stage, and introduce the temporal profile to mimic the evolution of the short lived strong magnetic field. In our approach, the spectra of the photon are studied in different scenarios in Sec.\ref{p5a}, and we find the yield of the photon is sensitive to the central classes as well as the evolution of magnetic field.
We use the result at $b=8$fm/c as an approximation of $20\%-40\%$ central class collisions, see in Fig.\ref{cdms}, it is found that the result  agrees with the measurements well. It is also found that in low $q_{T}$ region, the contribution on spectrum from gluon induced processes is small, but the enhancement becomes sizable in larger $q_{T}$ part: by means of about 20\% enhancement at $q_{T}=3$GeV, and even higher in larger $q_{T}$ region.
Through both qualitative and full numerical analysis in Sec.\ref{sec4.3}, we find collective flows of the photon has non-trivial behaviors due to glasma effect, which is different from \cite{Ayala:2017vex,Ayala:2019jey}. Even through, the photons have limited enhancement on total spectrum, the contribution on $v_{2}$ is obvious. We weight our results on top of PHSD model and find that when including the photon produced in magnetized glasma the result agrees experiment measure even better, e.g. from 0.0356 (only PHSD prediction \cite{Linnyk:2015tha}) to 0.1251 (weighted result) when comparing to measurement 0.1412$\pm$0.04 in \cite{PHENIX:2015igl} at $q_{T}=3$GeV by Conversion method. Besides, $v_{3}$ from our computation is finite and does not affect PHSD result. Additionally, we also provide the right tendency in $q_{T}>3\rm GeV$ region.

In application of MV model, we remove the IR divergence by a soft cut off  and UV divergence by the lattice spacing, such similar procedures have been widely carried out in many studies, see \cite{Lappi:2006hq,LappiWilson,Schenke:2012wb,FUKUSHIMA2012108,Schenke:2012hg} and so on. The evolving glasma is insensitive to the UV regulator, on the other hand, IR regulator affects the initial gluon multiplicity. By changing the IR regulator in our computation, the behaviors of collective flows are unaffected, but the spectra of the photon will be enhanced or reduced slightly. To obtain the same result, we can match $\tau_B$ technically, who plays as a parameter in our computation. Both the IR regulator and life time of the magnetic field have been reasonably chosen in our computation: $\Lambda = 150\rm MeV$ and $\tau_B = 0.05\rm fm/c$. 

Through the calculation presented in this work, the glasma effect in photon production via gluon fusion and splitting processes in strong magnetic field is  studied systematically, and result of improvements on PHSD prediction toward the measurements is discussed meticulously. Meanwhile, to improve the calculation in this work, contributions from high Landau-levels should be collected to enhance the solidness of current prediction. Besides, a rigorous treatment of the IR regulation triggers another attractive topic in application of glasma in similar studies. And it is also possible to relax the boost invariant assumption by using a 3+1D calculation with a colored-particle-in-cell (CPIC) model \cite{Gelfand:2016yho,Ipp:2017lho}. As our future efforts, these topics will be reported elsewhere soon.

\section*{Acknowledgments}
The authors acknowledge Xin-Li Sheng for inspiring discussion and Marco Ruggieri for valuable comments. This work is supported in part by the National Natural Science Foundation of China (NSFC) (Grant Nos: 11735007, 11890711, 12275104), and by the National Key Research and Development Program of China (No. 2022YFA1605501).

\bibliography{BIBL}

\end{document}